\documentclass[a4paper,fleqn]{article} 
\usepackage{modsim}
\usepackage{times}
\usepackage{natbib} 
\usepackage{amsmath, amssymb, amsthm} 

\MODSIMhead{X. Luo and P. Shevchenko, When to Bite the Bullet? - Optimal Strategies  for Reducing Global Warming} 

\usepackage{rotating}
\usepackage{amsbsy,enumerate}
\usepackage{graphicx}
\usepackage{ccaption}

\makeatletter
\renewcommand{\fnum@figure}[1]{\textbf{\figurename~\thefigure}. }
\renewcommand{\fnum@table}[1]{\textbf{\tablename~\thetable}. }
\makeatother

\begin{document}

\title{When to Bite the Bullet? - A Study of Optimal Strategies for Reducing Global Warming}

\author{\underline{X. Luo}, P. V. Shevchenko \address[A1]{\it{CSIRO Computational Informatics, Sydney NSW Australia}}
{}} 

\email{Xiaolin.Luo@csiro.au} 

\date{March 2009}


\begin{abstract}
   This work is based on the framework proposed by
\citet{Conrad:1997} to determine the optimal timing of an investment
or policy to slow global warming. While Conrad formulated the
problem as a stopping rule option pricing model, we treat the policy
decision by considering the total damage function that enables us to
make some interesting extensions to the original formulation. We
show that Conrad's framework is equivalent to minmization of the
expected value of the damage function under the stochastic optimal
stopping rule. We extend Conrad's model by allowing for policy cost
to grow with time. In addition to closed form solution, we also
perform Monte Carlo simulations to find the distribution for the
total damage and show that at higher quantiles the damage may become
too large and so is the risk on the global economy. We also show
that the decision to take action largely depends on the cost of the
action.  For example, in the case of model parameters calibrated as
in \citet{Conrad:1997} with a constant cost, there is a rather long
wait before the action is expected to be taken, but if the cost
increases with the same rate as the global economy growth, then
action has to be taken immediately to minimize the damage.

\end{abstract} 

\maketitle

\section{INTRODUCTION}

Over the last few decades, scientists have studied extensively the
effect of global warming and optimal strategies to slow down the
process to manage disaster risks. For example, see
\citet{Nordhaus1991}, \citet{Nordhaus1992}, \citet{Nordhaus2000},
\citet{AAG2003}, \citet{UNFCCC2006}, \citet{Nordhaus2008},
\citet{Greiner2010}, \citet{IPCC2012}. Some of these models
establish the close linkage between global economy growth and the
damage by global warming and support the view that policy actions
against global warming are urgently needed.

 In this paper, we study the global warming
model of \citet{Conrad:1997} and suggest optimal timing for the
policy by considering the total damage function. The global
temperature is assumed to be drifting upward following a Brownian
motion. Investments to slow the global warming are called ``bullets"
and they are characterized by the reduction of both the drift rate
and the standard deviation  after the investments. If the action is
taken at time $t=\tau$ which reduces the drift from $\mu_1$ to
$\mu_2$ and the volatility from $\xi_1$ to $\xi_2$, the temperature
process is assumed to be
\begin{equation}
\label{eq_temp} dC_t=\mu(t)dt+\xi(t)dz, \quad \mu(t)=\mu_1
I_{t<\tau}+\mu_2 I_{t\ge \tau}, \quad \xi(t)=\xi_1 I_{t<\tau}+\xi_2
I_{t\ge \tau},
\end{equation}

where $C_t$ is the mean global temperature at time $t$, $\mu(t)$ is
the drift, $\xi(t)$ is the volatility, $dz$ is the increment of the
standard Wiener process and $I_{\{.\}} $ is the indicator symbol
that equals 1 if the condition in \{.\} is true and 0 otherwise. The
damage rate from global warming is assumed to be the convex function
\begin{equation}
\label{eq_damage_temp} S_t=S_0 e^{\gamma(C_t-C_0)},
\end{equation}
where $S_t$ is the damage rate in billions of dollars per annum,
 $\gamma$ is a positive parameter,  $C_0$ is the reference
temperature and $S_0$ is the corresponding damage rate. The original
model was calibrated using time series data for temperature
anomalies and estimates of the damage from the global warming and
the cost of the investments. In this study we extend Conrad's model
by allowing for policy cost to grow with time. In addition to closed
form solution for the optimal timing of the policy, we also perform
Monte Carlo simulations to find the distribution for the possible
total damage.

\section{STOPPING-RULE MODEL AND SOLUTION}

\label{sec:stop} Applying Ito's Lemma to function
(\ref{eq_damage_temp}) with process (\ref{eq_temp}), one can show
that the damage
 rate $S_t$ from global warming follows the  geometric Brownian
 motion with dirft $\alpha(t)$ and volatility $\sigma(t)$
\begin{equation}
\label{eq_damage} dS_t=\alpha(t) S_t dt + \sigma(t) S_t dz,\quad
\alpha(t)=\alpha_1 I_{t<\tau}+\alpha_2 I_{t\ge \tau},\quad
\sigma(t)=\sigma_1 I_{t<\tau}+\sigma_2 I_{t\ge \tau},
\end{equation}
where  $\alpha_1=\mu_1\gamma+(\xi_1\gamma)^2/2$ and
$\sigma_1=\xi_1\gamma$ are the drift and volatility before the
policy is introduced,
$\alpha_2=\mu_2\gamma+(\xi_2\gamma)^2/2<\alpha_1$ and
$\sigma_2=\xi_2\gamma<\sigma_1$ are the drift and volatility after
the policy, $dz$ is the increment of a standard Wiener proces. We
assume that the cost of the policy changes in time as
$K_t=K\exp(q\times t)$ and it is paid only once when the policy is
introduced at time $t=\tau$. It is important to note that
\citet{Conrad:1997} assumes that the cost of policy is time
independent. It is further assumed that the global economy grows at
a rate $r> \alpha_1$.

From (\ref{eq_damage}) it follows that the  damage rate is
log-normally distributed, thus the expected value of the damage at
future time $T$, given damage rate $S_t$ at time $t$,  is
$E_t[S_T]=S_t\exp(\int_t^T\alpha(x)dx)$. Hereafter, notation
$E_t[\cdot]$ denotes expectation conditional on information
available at time $t$. Thus the present value at time $t$ of the
future total expected damage if the policy is never introduced can
be easily calculated as
\begin{equation}
\label{eq_int_damage1} d_1(S_t)=E_t \left[ \int_t^\infty
S_{t^\prime} e^{-r(t^\prime-t)} dt^\prime \right] = S_t\int_t^\infty
e^{\alpha_1(t^\prime-t)} e^{-r(t^\prime-t)} dt^\prime
=\frac{S_t}{(r-\alpha_1)}.
\end{equation}
  Similarly, if the policy is introduced at time $t$, the total discounted
expected future damage plus the policy cost is given by
\begin{equation}
d_2(S_t)=E_t \left[ \int_t^\infty S_{t^\prime}e^{-r(t^\prime-t)}
dt^\prime \right]+K_t = \frac{S_t}{(r-\alpha_2)}+K_t.
\end{equation}
 Consider the actual total damage function represented by random
variable
\begin{equation}\label{total_damage_eq}
D(t,\tau) =\int_t^\tau S_x e^{-r(x-t)} dx+\int_\tau^\infty
S_xe^{-r(x-t)} dx + K_\tau e^{-r(\tau-t)}.
\end{equation}
Here, $\tau$ is stochastic stopping time, i.e. different
realizations will have different $\tau$. The expected total damage
under the stopping rule $\tau$ for damage rate $S$ at time $t$ is
\begin{equation}
V(S,t)=E_t\left[D(t,\tau) \right].
\end{equation}
Typically, one defines an optimal stopping rule $\tau^\ast$ as the
one that minimizes expected value of $D(t,\tau)$, i.e. the expected
total damage under the optimal stopping rule is
\begin{equation}
V^\ast(S,t) =\min _\tau E_t\left[D(t,\tau)
\right]=E_t\left[D(t,\tau^\ast) \right].
\end{equation}
We can solve this problem by formulating partial differential
equation (PDE) for $V(S,t)$; for detailed discussion of PDE approach
for valuation of projects and investment rules, see
\citet{DixitPindyck1994}.

It is interesting to consider the deterministic case when the
volatility $\sigma_1=\sigma_2=0$ in (\ref{eq_damage}). In this case
$S_t=S_0e^{\alpha_1t}$ when $t<\tau$ and $S_t=S_0e^{\alpha_1\tau+
\alpha_2(t-\tau)}$ when $t\geq\tau$.   Minimizing  the deterministic
function $D(t,\tau)$ with respect to $\tau$ we obtain the optimal
time $\tau^* $ and the corresponding critical damage rate $S^*$
\begin{equation}
\label{eq_determ} \tau^* =\frac{1}{\alpha_1-q}\ln \left (
\frac{K(r-q)(\alpha_2-r)}{S_0(\alpha_2-\alpha_1)} \right),  \;\;\;
S^*=S_0\left ( \frac{K(r-q)(\alpha_2-r)}{S_0(\alpha_2-\alpha_1)}
\right)^{\frac{\alpha_1}{\alpha_1-q}},
\end{equation}
 if $q<\alpha_1$. If the process is stochastic instead of deterministic,
 there is some value in waiting somewhat longer than the
 deterministic case, as will be shown later with an actual example.

Now for the stochastic process consider the region where the policy
is not introduced, i.e. the transition density function
$f(S^{\prime},t^\prime|S,t)$ for the damage rate stochastic process
(\ref{eq_damage}) satisfies the backward Kolmogorov equation
\begin{equation}
\frac{\partial f(S^{\prime},t^\prime|S,t)}{\partial t}+\alpha_1
S\frac{\partial f(S^{\prime},t^\prime|S,t)}{\partial
S}+\frac{1}{2}\sigma_1^2 S^2 \frac{\partial^2
f(S^{\prime},t^\prime|S,t)}{\partial S^2}=0;
\end{equation}
for transition densities and Kolmogorov equations corresponding to
Wiener processes, see \citet{CoxMiller1965}. Multiplying each term
by $S^\prime e^{-r(t^\prime-t)}$, integrating from $t$ to $\infty$
and taking expectation yields

\begin{equation}\label{PDE_totaldamage_eq}
\frac{\partial V(S,t)}{\partial t}+\alpha_1 S\frac{\partial
V(S,t)}{\partial S}+\frac{1}{2}\sigma_1^2 S^2 \frac{\partial^2
V(S,t)}{\partial S^2}-rV(S,t)+S=0,
\end{equation}
 subject to the condition $V(S,t)\le d_2(S)+K_t$, where the equality applies on the boundary $S=H_t$ when policy is
 introduced.

Note that the terms $rV$ and $S$ appear because we move
$e^{-r(t^\prime-t)}$ and integration $\int_t^\infty$ under the
derivative $\partial f/\partial t$. Also note that we do not start
the policy if $V(S,t)<d_2(S(t))$ and introduce the policy
 otherwise.

 The policy cost is time dependent $K_t=K\exp(q\times t)$,
thus the problem is time non-homogenous. Assume that the policy is
introduced when damage rate $S_t$ breaches the level
$H_t=H\exp(q\times t)$. To reduce the problem to time homogeneous
and solve PDE, consider a new variable $Y=S\exp(-q\times t)$. The
PDE for $Q(Y,t)\equiv V(S,t)$ will be the same as
(\ref{PDE_totaldamage_eq}) except the change of $\alpha_1$ to
$\alpha_1-q$. Substituting solution in the form
$Q(Y,t)=\widetilde{Q}(Y)\exp(q\times t)$ leads to ODE for
$\widetilde{Q}(Y)$
\begin{equation}\label{ODE_totaldamage_eq}
(\alpha_1-q) Y \frac{d\widetilde{Q}(Y)}{dY} + \frac{\sigma_1^2}{2}
Y^2 \frac{d^2\widetilde{Q}(Y)}{dY^2}-(r-q)\widetilde{Q}(Y)+Y=0,
\end{equation}
subject to the condition $ \widetilde{Q}(Y)\le
\frac{Y}{(r-\alpha_2)}+K$, where the equality applies to the
boundary at $Y=H$. Equation (\ref{ODE_totaldamage_eq}) is a
non-homogeneous, second-order differential equation. The homogeneous
part has the well-known solution $\widetilde{Q}(Y)=\xi
Y^{\tilde{\epsilon}} + \eta Y^{\epsilon}$, where $\tilde{\epsilon}$
and $\epsilon$ are given by
\begin{eqnarray}
\label{eq_root} \tilde\epsilon &=&(1/2-(\alpha_1-q)/\sigma_1^2) -
\sqrt{(1/2-(\alpha_1-q)/\sigma_1^2)^2+2(r-q)/\sigma_1^2 };\\
\epsilon &=&(1/2-(\alpha_1-q)/\sigma_1^2) +
\sqrt{(1/2-(\alpha_1-q)/\sigma_1^2)^2+2(r-q)/\sigma_1^2 },
\end{eqnarray}
and constants $\xi$, $\eta$ can be found from the boundary
conditions.  One of the boundary conditions is $\widetilde{Q}(0)=0$.
It can be shown that if $q<r$ then $\tilde{\epsilon} <0$.   Thus the
conditions $\widetilde{Q}(0)=0$ and $\tilde{\epsilon} <0$ lead to
$\xi=0$ and solution simplifies to $\widetilde{Q}(Y)=\eta
Y^{\epsilon}$. On the other hand, if $q\geq r$, $\tilde{\epsilon}
\geq 0$, both terms $\xi Y^{\tilde{\epsilon}} $ and $\eta
Y^{\epsilon}$ are retained in the solution. Here we only consider
the solution for  $q < r$.

 A particular solution to equation (\ref{ODE_totaldamage_eq}) is simply
${Y}/{(r-\alpha_1)}$. Thus a general solution is obtained by adding
the solution of the homogenuous portion to the particular solution
yielding
\begin{equation}
\widetilde{Q}(Y) =\eta Y^{\epsilon} +\frac{Y}{(r-\alpha_1)}.
\end{equation}
The continuity conditions for the function on the boundary $Y=H$,
i.e. $\widetilde{Q}(Y) = \frac{Y}{(r-\alpha_2)}+K$, gives
\begin{equation}
 \eta H^{\epsilon}
+\frac{H}{(r-\alpha_1)}=\frac{H}{(r-\alpha_2)} +K,
\end{equation}
that allows to identify constant $\eta$
\begin{equation}
\eta=\frac{1}{H^{\epsilon-1}}\frac{(\alpha_2-\alpha_1)}{(r-\alpha_1)(r-\alpha_2)}+\frac{K}{H^\epsilon}.
\end{equation}
Thus the solution for total damage is $V(S,t)=\widetilde{Q}(S
e^{-qt})e^{qt}$, where
\begin{eqnarray}
 \widetilde{Q}(Y)&=&\eta
Y^{\epsilon} +\frac{Y}{(r-\alpha_1)}=
\left(\frac{K}{H^\epsilon}-\frac{1}{H^{\epsilon-1}}\frac{(\alpha_1-\alpha_2)}{(r-\alpha_1)(r-\alpha_2)}\right)
Y^{\epsilon}
+\frac{Y}{(r-\alpha_1)}\nonumber\\
&=&\left(\frac{Y}{H}\right)^\epsilon
\left(K-H\frac{\alpha_1-\alpha_2}{(r-\alpha_1)(r-\alpha_2)}\right)+\frac{Y}{(r-\alpha_1)}.
\end{eqnarray}
Considering $V(S,t)$ as a function of boundary level $H$, we can
find optimal level $H^\ast$, where $V(S,t)$ is minimized with
respect to $H$ by solving
\begin{equation}
\frac{\partial \widetilde{Q}(Y)}{\partial
H}=Y^\epsilon\left(-\epsilon\frac{K}{H^{\epsilon+1}}+\frac{\epsilon-1}{H^\epsilon}\frac{\alpha_1-\alpha_2}
{(r-\alpha_1)(r-\alpha_2)}\right)=0,
\end{equation}
that gives
\begin{equation}
H^* =\frac{\epsilon
(r-\alpha_1)(r-\alpha_2)K}{(\epsilon-1)(\alpha_1-\alpha_2)}.
\end{equation}
We could obtain the same result if we would solve PDE
(\ref{PDE_totaldamage_eq}) for optimal $V^\ast(S,t)$ directly; in
that case we just had to impose continuity condition not just for
the function but also for its 1st derivative on the stopping
boundary; for discussion about these boundary conditions, see
\citet[pp. 130-132]{DixitPindyck1994}.

The corresponding optimal critical damage rate is $S^\ast=
e^{qt}H^\ast$. The probability density of the hitting time $\tau$
for process $Y_t$ to breach level $H$ (i.e. $S_t$ to breach $H_t$)
is given by
\begin{equation}\label{eq_pdf_hit}
g(\tau)=\frac{\ln(H/S_0)}{\sigma_1 \tau\sqrt{2\pi
\tau}}\exp\left\{-\frac{[\ln(H/S_0)-\eta \tau]^2}{2\sigma_1^2\tau
}\right\}, \quad \eta=\alpha_1-q-\sigma_1^2/2>0, \;\;\text{for}\;
H>Y_0=S_0.
\end{equation}
The above density is only valid if $\eta>0$. From (\ref{eq_pdf_hit})
we find the expected hitting time is
\begin{equation}\label{eq_hit_time}
\text{E}[\tau]= \frac{ \ln(H/S_0) }{\eta}.
\end{equation}
In the case of time independent cost (i.e. $q=0$), the expected
critical time (to breach the optimal critical level $S^\ast$)
 from the above equation reduces to
$(C^\ast-C_0)/\mu_1$, where $C^\ast=C_0+\ln(S^\ast/S_0)/\gamma$ is
the critical temperature corresponding to $S^\ast$. The solution
obtained here for $q=0$ is equivalent to the solution in
\citet{Conrad:1997}. Finally, the solution to the stochastic problem
also contains the deterministic case given in (\ref{eq_determ}): by
letting $\sigma_1=\sigma_2\rightarrow 0$, we find
$\epsilon\rightarrow (r-q)/(\alpha_1-q)$ and the solution for
$\tau^\ast$ and $S^\ast$ is identical to (\ref{eq_determ}).

\section{NUMERICAL RESULTS}
Here we analyse the distribution of total damage $D(t,\tau)$ given
by (\ref{total_damage_eq}) under the optimal and non optimal
stopping rules. Monte Carlo simulation of $D(t,\tau)$ is simple: one
has to simulate damage rate process $S(t)$ given by
(\ref{eq_damage}) with drift $\alpha_1$ until time $\tau$ where
$S(t)$ breaches the critical level $H \exp(q\times t)$ and with
drift $\alpha_2$ afterwards, and calculate $D(t,\tau)$ in
(\ref{total_damage_eq}). Repeating simulation many times allows us
to find a distribution of $D(t,\tau)$.

The parameter values of $r=0.05$, $\alpha_1=0.03708657$,
$\alpha_2=0.02758027$, $\sigma_1=\sigma_2=0.19012608$ are taken from
\citet{Conrad:1997}. For the base solution, we assume a constant
cost, i.e. $q=0.0$, that corresponds to Conrad's work. The case of
constant cost assumes the future value of the cost remains the same,
which effectively make the present value of the cost getting cheaper
at the same rate as the economy growth rate, a somewhat unrealistic
assumption.

 Typically, the solution to the stopping rule problem, or
equivalently to the American option pricing problem, is in terms of
expected value of payoff function. In the context of financial
options, this expectation gives a fair value at which there are
buyers and sellers willing to do the transaction, knowing that the
associated risks can be fully hedged. If these kind of buying and
selling are repeated many times, on average these transactions will
give the buyers or sellers a fair return adjusted to the risks they
take. However, to the question of when to take  action to slow down
global climate change, the answer based on average damage (expected
value) is too risky and almost does not make sense. There is only
one realization of reality and one chance for us to do it right. The
risk to human being's survival is too great for us to rely on an
average solution to manage the risk.

Indeed, the solution based on the expectations gives $\tau^*=122$
years as the expected optimal time to take action, a seemingly very
long time. The optimal critical damage rate $S^*=\text{USD }10.19$
billion per annum at which the action should be taken to minimize
the expected total damage. The corresponding critical temperature is
$C^* =16.22$ degrees, assuming today's temperature is $C_0 =15$
degrees. Using 100,000 Monte Carlo simulations, the expected total
damage from this optimal strategy is about \text{USD } 61.6 billion
with a numerical standard error of 0.83 billion due to finite number
of simulations.

In comparison with the  the above solution for the stochastic case,
the corresponding solution for the deterministic case  in
(\ref{eq_determ}) gives $\tau^*=53$ years and $S^*=\text{USD } 7.08$
billion per annum, both are much smaller than the stochastic
counterparts, demonstrating the value in waiting under uncertainty.

\begin{figure}
\begin{center}
\includegraphics[width=12cm, height=7.cm]{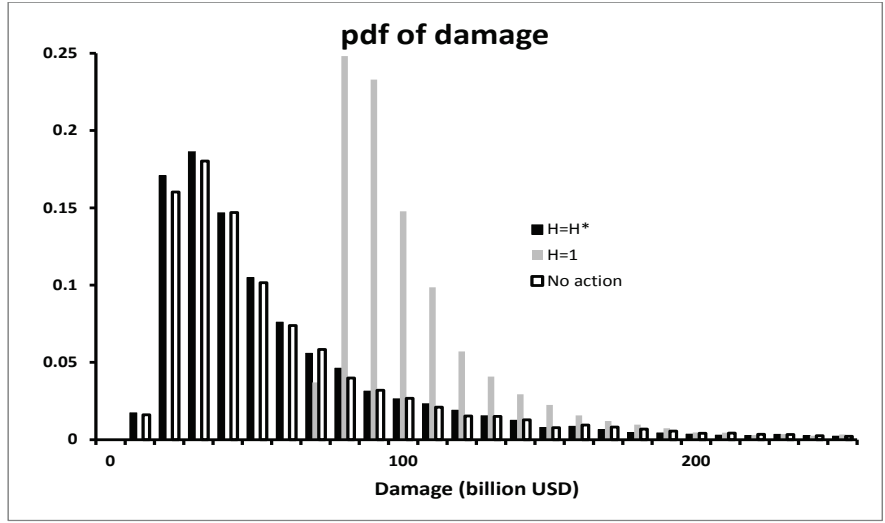}
\caption{Probability density function (pdf) for the
 total damage. $H=1$ is for taking action
immediately, $H=H^\ast$ is for optimal strategy and ``No action"
corresponds to never taking any action (business as usual). }
\end{center}
\end{figure}

\begin{figure}
\begin{center}
\includegraphics[width=12cm, height=7cm]{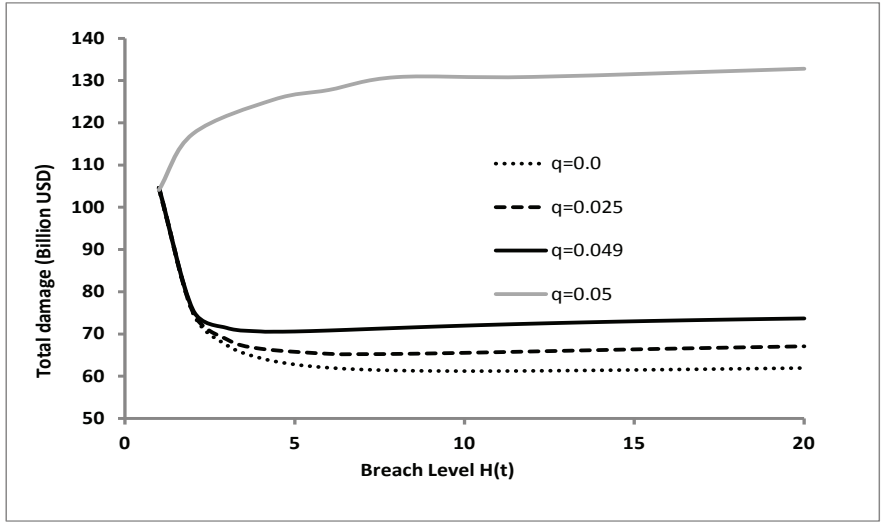}
\caption{Damage curves for different cost increasing rate $q$. The
curve for $q=0.05$ is from Monte Carlo simulation, and all others
are analytical solutions.}
\end{center}
\end{figure}

Instead of considering  the expectations, a more sensible approach
is by considering the entire distribution of the total damage. The
Monte Carlo simulation allows us to obtain such a distribution. The
pdf of total damage for the case of constant cost is shown in Figure
1. Given the distribution, higher quantiles can be considered for
decision making with regard to managing risks, similar to the
concept of Value at Risk used in managing financial risks. If the
optimal critical total damage level of \text{USD } 61.6 billion is
just acceptable to the global environment and economy, there is
about $30\%$ chance this total damage will be exceeded and
disastrous consequences may result. The $90\%$ quantile of the total
damage is $\text{USD } 121.5$ billion, twice as large as the
expected value. In other words, there is $10\%$ chance the total
damage could reach twice as high as the acceptable level, which may
be a disastrous outcome if this strategy is taken.

To see the impact on non-optimal stopping rule, we find
distributions of $D(t,\tau)$ when stopping rule corresponds to
 different breaching levels $H=\omega S^\ast$ with $\omega$ ranging around 1.
For example, at $H=2.0$, the expected total damage obtained from
Monte Carlo simulation is about 74. The corresponding critical
temperature at which the action is taken is now 15.36 degrees, and
the corresponding expected waiting time is 36 years. Comparing with
the optimal strategy, this much earlier action at a much lower
critical level of damage rate causes more total damage! At $H=1.0$,
i.e. if we bite the bullet now without any waiting,  the expected
total damage obtained from Monte Carlo simulation is 104.2 (the
exact value is 104.6), which is even larger than the damage of
waiting for 36 years with $H=2.0$.  Such is the paradox of the
current model with these specific inputs. Finally, if no action is
taken (the bullet is never bit), the expected total damage is 77.4
which is even better than taking action immediately. This is because
the reduction in the damage rate due to immediate action is not
large enough to justify the cost. The pdf curves of damage for the
cases of no action and immediate action are also shown in Figure 1.
The corresponding quantiles from Monte Carlo simulations for these
three distributions are shown in Table 1.

\begin{table} \label{table:one}
\caption{Quantiles of total damage (billion USD)}
\begin{center}
\begin{tabular}{cccc}
\hline
Cases & Mean & Median  & 0.9 quantile \\
\hline
No action & 75.8 & 39.8 & 140.2 \\
Immediate action& 104.2& 89.1 & 143.7\\
Optimal strategy & 61.6 & 39.3 & 121.5\\

\hline
\end{tabular}
\end{center}
\end{table}%

 In general, under the above  inputs, any earlier (later)
action or equivalently action at a lower (higher) threshold of the
damage rate will be worse off than the optimal strategy.
Intuitively, this does not seem to be expected: an earlier
 action does not reduce the total damage of global warming. This
 paradox can be explained by comparing the cost with the accumulated
 damage. The assumption of constant future cost means the present
 value of the cost decreases exponentially with time at the discount
 rate, and if this decrease is faster than the increase in the total
 damage in absolute terms, it is worthwhile to wait for some time before taking the
  action when the cost is much cheaper.

  In the scenarios considered above, even the optimal strategy may
  not be good enough because of the high risk of too large a damage to
  the global economy. Assuming the process and especially the damage are correctly modeled,
 the only way out for confidently saving the global economy is to
 take more drastic measures to reduce the damage rate more
 aggressively with reduced cost. This means much reduced drift rate $\alpha_2$
 after the bullet is bit, and at the same time keep the cost $K_t$ sufficiently low to enable
  effective and confident reduction of the total damage under the optimal
  strategy. This can be achieved through scientific research and technology breakthroughs.

The above paradox of earlier action being worse off than waiting
disappears when we consider a more realistic scenario for the cost.
Assuming the future cost increases with the discount rate, i.e.
$q=r$, we then obtain a qualitatively different results. In this
case, Monte Carlo simulation shows that the total damage becomes a
monotonically increasing function of the breaching level. That is,
the higher the critical value we take as the trigger of action, the
larger the total damage will be. There is no optimal level in any
delayed actions. The longer we wait, the larger will be the total
damage. In this scenario the best approach is to take the action
immediately.

For other cost increment rates $0<q<r$, the damage monotonically and
continuously increases with the value of $q$. Figure 2 shows the
damage curves corresponding to different values of $q$. For $q<r$
there is a minimum corresponding to the optimal policy.   As the
cost increasing rate $q$ approaches the economic growth rate
$r=0.05$ from below, the damage curve converges to an asymptotic
line. At $q\geq r$ the solution jumps to a different branch with
optimal solution corresponding to an immediate action.  This jump is
easily confirmed by Monte Carlo simulation, see Fig. 2.

\section{CONCLUSIONS}
\label{sec:conclusionsomparison} Finding optimal stopping rule via
PDE approach is a very powerful method. In this paper the solution
is found in closed form for $0<q<r$. Qualitatively, the optimal time
to bite the bullet depends on, among other factors,  the relative
rates of economic growth and cost increment. For more realistic
stochastic process for the damage rate and policy time dependence
different from exponential, PDE can be solved numerically which is a
subject of our future research.

\end{document}